\documentclass[conference]{IEEEtran}
\IEEEoverridecommandlockouts
\usepackage[utf8]{inputenc}
\usepackage[T1]{fontenc}
\usepackage[english]{babel}
\usepackage{cite}
\usepackage{amsmath,amssymb,amsfonts,amsthm}
\usepackage{graphicx}
\usepackage{textcomp}
\usepackage{xcolor}
\usepackage{booktabs}
\usepackage{algorithm}
\usepackage{algorithmicx}
\usepackage{todonotes}
\usepackage[autostyle]{csquotes}
\usepackage[printonlyused]{acronym}
\usepackage{multirow}
\usepackage{xspace}
\usepackage{ctable}
\usepackage{caption}
\usepackage{relsize}
\usepackage{subcaption}
\usepackage[colorlinks=true,citecolor=blue,filecolor=blue,linkcolor=red,urlcolor=red]{hyperref}
\usepackage[inline]{enumitem}
    \setlist{nosep}

\usepackage[margin=0.7in]{geometry}
\usepackage[parfill]{parskip}
\usepackage{pgfplots}
\pgfplotsset{compat=1.14}
\usepackage{adjustbox}
\usepackage{scalefnt}
\usetikzlibrary{calc}
\usepackage{setspace}

\usepackage{soul}

\addto\extrasenglish{%
}

\newacro{AST}{abstract syntax tree}
\newacro{DP}{dynamic programming}
\newacro{HLS}{high-level synthesis}
\newacro{SPMD}{single program, multiple data}
\newacro{GCUPS}{giga cell updates per second}

\def\CC{C\nolinebreak[4]\hspace{-.05em}\raisebox{.4ex}{\relsize{-3}{\textbf{++}}}\xspace}

\def\impala{Impala\xspace}

\usepackage{listings}
\usepackage{lstautogobble}
\usepackage{etoolbox}

\expandafter\patchcmd\csname \string\lstinline\endcsname{%
    \leavevmode
    \bgroup
}{%
    \leavevmode
    \ifmmode\hbox\fi
    \bgroup
}{}{%
    \typeout{Patching of \string\lstinline\space failed!}%
}

    {\noindent\ignorespaces\begin{lstlisting}[#1,float,floatplacement=H]}{\end{lstlisting}\noindent\ignorespacesafterend}

\definecolor{whitesmoke}{rgb}{0.96, 0.96, 0.96}

\definecolor{lightblue}{rgb}{0.68, 0.85, 0.9}
\definecolor{darkspringgreen}{rgb}{0.09, 0.45, 0.27}

\definecolor[named]{codegreen}    {named}{darkspringgreen}
\definecolor[named]{codered}      {named}{red}
\definecolor[named]{codelightblue}{named}{lightblue}
\definecolor[named]{codedarkblue} {named}{blue}

\def\lst{\lstinline}

\makeatletter
\newenvironment{btHighlight}[1][]
{\begingroup\tikzset{bt@Highlight@par/.style={#1}}\begin{lrbox}{\@tempboxa}}
{\end{lrbox}\bt@HL@box[bt@Highlight@par]{\@tempboxa}\endgroup}

\newcommand\btHL[1][]{%
    \begin{btHighlight}[#1]\bgroup\aftergroup\bt@HL@endenv%
}
\def\bt@HL@endenv{%
    \end{btHighlight}%
    \egroup
}
\newcommand{\bt@HL@box}[2][]{%
    \tikz[#1]{%
        \pgfpathrectangle{\pgfpoint{1pt}{0pt}}{\pgfpoint{\wd #2}{\ht #2}}%
        \pgfusepath{use as bounding box}%
        \node[anchor=base west, fill=codelightblue,outer sep=0pt,inner xsep=1pt, inner ysep=0pt, rounded corners=3pt, minimum height=\ht\strutbox+1pt,#1]{\raisebox{1pt}{\strut}\strut\usebox{#2}};
    }%
}
\makeatother

\lstdefinestyle{node}{
    backgroundcolor=,
    language=,
    basicstyle=\tiny\ttfamily,
    morekeywords = {br,neg,or,and,all,any},
    numbers=none,
    mathescape=true,
    frame=none,
    literate={<-}{{$\leftarrow$}}1
}

\lstdefinelanguage{alg}{
    morecomment = [s]{/*}{*/},
    morecomment = [l]{//},
    sensitive = true,
    morekeywords = {for,next,to,step}
}

\lstdefinelanguage{impala}{
    morecomment = [s]{/*}{*/},
    morecomment = [l]{//},
    sensitive = true,
    morekeywords = {i8,i16,i32,i64,u8,u16,u32,u64,f16,f32,f64,bool,int,float,double,extern,struct,as,match,true,false,type,with,let,mut,while,in,exit,return,break,continue,if,else,for,do,fn,enum},
    moredelim=**[is][\btHL]{§}{§},
    morestring=[b]",
}

\lstdefinelanguage{metaocaml}{
    sensitive = true,
    morekeywords = {let,in,rec,if,then,else,fun},
    moredelim=**[is][\btHL]{§}{§},
}

\lstdefinelanguage{pseudoml}{
    sensitive = true,
    morekeywords={fun,where,whererec,lambda,let,letrec,in,and,bool,float,int,br,noret},
    literate=%
        {==}{{=}}1
        {!=}{{$\neq$}}1
        {<=}{{$\leq$}}1
        {>=}{{$\geq$}}1
        {->}{{$\rightarrow$}}1
        {<-}{{$\leftarrow$}}1
        {bot}{{$\bot$}}1
        {LAMBDA}{{$\lambda$}}1
}

\lstdefinelanguage{scala}{
    morecomment = [s]{/*}{*/},
    morecomment = [l]{//},
    sensitive = true,
    morekeywords = {val,var,new,with,import,trait,this,def,if,else,Int},
    moredelim=**[is][\btHL]{§}{§},
}

\lstdefinelanguage{scheme}{
    sensitive = true,
    morekeywords={define,filter}
}

\lstdefinelanguage{sierra}{
    morecomment = [s]{/*}{*/},
    morecomment = [l]{//},
    morestring=[b]",
    sensitive = true,
    morekeywords = {uniform,varying,simd,scalar,for_each_active,for_each_unique,current_mask},
    morekeywords = {kernel,uint,mask,skip,true,false,uint32_t,uint64_t,nullptr,return,public,protected,private,template,auto,class,virtual,struct,union,void,this,size_t,volatile,if,else,do,while,case,goto,switch,for,while,bool,typedef,static,const,float,int,short,char,double,break,continue},
    keywords = {[2]define},
    keywordstyle={[2]\color{uds-purple}\bfseries},
    moredelim=**[is][\btHL]{§}{§},
}

\lstdefinelanguage{ssa}{
    sensitive = true,
    morekeywords={fn,bool,float,int,phi,goto,br,return},
    literate=
        {:=}{{$\gets$}}1
        {==}{{=}}1
        {!=}{{$\neq$}}1
        {<=}{{$\leq$}}1
        {>=}{{$\geq$}}1
        {->}{{$\rightarrow$}}1
        {<-}{{$\leftarrow$}}1
        {PHI}{{$\phi$}}1
}

\lstdefinelanguage{terra}{
    morecomment = [s]{/*}{*/},
    morecomment = [l]{//},
    sensitive = true,
    morekeywords = {int,function,if,then,return,else,elseif,terra,end,local},
    moredelim=**[is][\btHL]{§}{§},
}

\newcommand{\CopyrightNotice}{\hbox%
    {\parbox{18cm}{\textsf\centering\scriptsize$ $\\[-14cm]
    \centering\setstretch{.3}%
    \textcopyright~2020 IEEE.
    Personal use of this material is permitted.
    Permission from IEEE must be obtained for all other uses, in any current or future media, including reprinting/republishing this material for advertising or promotional purposes,creating new collective works, for resale or redistribution to servers or lists, or reuse of any copyrighted component of this work in other works.\par}%
    \vspace{-\baselineskip}}%
}

\begin{document}

\title{AnySeq: A High Performance Sequence Alignment Library based on Partial Evaluation\\
\thanks{This work is supported by the Federal Ministry of Education and Research (BMBF) as part of the MetaDL, Metacca, and ProThOS projects as well as by the Intel Visual Computing Institute (IVCI) and Cluster of Excellence on Multimodal Computing and Interaction (MMCI) at Saarland University.}
}

\author{%
    \IEEEauthorblockN{%
        Andr\'{e} M\"uller\IEEEauthorrefmark{1},
        Bertil Schmidt\IEEEauthorrefmark{1},
        Andreas Hildebrandt\IEEEauthorrefmark{1},
        \\
        Richard Membarth\IEEEauthorrefmark{2}\,\IEEEauthorrefmark{3},
        Roland Leißa\IEEEauthorrefmark{3},
        Matthis Kruse\IEEEauthorrefmark{3},
        Sebastian Hack\IEEEauthorrefmark{3}}%
    \IEEEauthorblockA{\IEEEauthorrefmark{1}%
            Johannes Gutenberg University,
            \{muelan,bertil.schmidt,andreas.hildebrandt\}@uni-mainz.de}%
    \IEEEauthorblockA{\IEEEauthorrefmark{2}%
        DFKI, Saarland Informatics Campus, richard.membarth@dfki.de}%
    \IEEEauthorblockA{\IEEEauthorrefmark{3}%
        Saarland University, Saarland Informatics Campus, \{leissa,matthis.kruse,hack\}@cs.uni-saarland.de}}


\maketitle
\CopyrightNotice

\begin{abstract}
Sequence alignments are fundamental to bioinformatics which has resulted in a variety of optimized implementations. Unfortunately, the vast majority of them are hand-tuned and specific to certain architectures and execution models. This not only makes them challenging to understand and extend, but also difficult to port to other platforms. We present AnySeq---a novel library for computing different types of pairwise alignments of DNA sequences. Our approach combines high performance with an intuitively understandable implementation, which is achieved through the concept of partial evaluation. Using the AnyDSL compiler framework, AnySeq enables the compilation of algorithmic variants that are highly optimized for specific usage scenarios and hardware targets with a single, uniform codebase. The resulting domain-specific library thus allows the variation of alignment parameters (such as alignment type, scoring scheme, and traceback vs.~plain score) by simple function composition rather than metaprogramming techniques which are often hard to understand. Our implementation supports multithreading and SIMD vectorization on CPUs, CUDA-enabled GPUs, and FPGAs. AnySeq is at most 7\% slower and in many cases faster (up to 12\%) than state-of-the art manually optimized alignment libraries on CPUs (SeqAn) and on GPUs (NVBio).
\end{abstract}

\section{Introduction}

Recent years have seen a tremendous increase in the volume of data generated in the life sciences, especially propelled by the rapid progress of next-generation sequencing (NGS) technologies. As a consequence, modern bioinformatics tools often require highly efficient implementations of core sequence analysis algorithms.

Given a pair of genomic sequences, a common operation in bioinformatics is to identify their similarity under a model of evolution which allows for certain sequence modifications. This leads to so-called {\em sequence alignments} that map characters across the sequences in an order-preserving way while potentially inserting gaps such that a mathematical model of their similarity is maximized. For pairwise alignment computation, the Smith-Waterman algorithm~\cite{sw}, the Needleman-Wunsch algorithm~\cite{nw}, and their variants are widely used. These compute an optimal local, global, or semi-global alignment of two sequences under a given scoring scheme by means of \ac{DP}. However, the associated time complexity proportional to the product of sequence lengths makes this approach a time consuming component of various bioinformatics workflows. As a consequence, these algorithms have been optimized on numerous architectures including CPUs~\cite{swaphils, hou2016aalign, misra2018performance}, GPUs~\cite{cudalign, liu2013cudasw++, korpar2013sw, de2016cudalign}, and FPGAs~\cite{oliver2005hyper, fpga, rucci2018swifold}. Unfortunately, the majority of implementations are hand-tuned and specific to certain architectures and execution models. This not only makes them challenging to understand and extend, but also difficult to port to other platforms. For a typical bioinformatics setting it would be more desirable to use a flexible library that can exploit a variety of modern hardware. This motivates the design of reusable and extensible sequence alignment components that can ensure compatibility and performance while at the same time reducing bioinformatics application development time. Existing optimized alignment libraries have so far only targeted specific architectures using either C/\CC for CPUs~\cite{reinert2017seqan, parasail, ssw} or CUDA~\CC for GPUs~\cite{nvbio}.

\subsection{Contributions}

We present AnySeq, a novel high-performance library for computing pairwise alignments of genomic sequences implemented using the AnyDSL compiler~\cite{DBLP:conf/gpce/LeissaBHMS15,DBLP:journals/pacmpl/LeissaBHPMSMS18}. AnyDSL and our approach is based upon the concept of {\em partial evaluation}~\cite{Futamura:1999:PEC:609149.609205,Consel,Brady}, which allows the compilation of different variants of the \ac{DP} algorithm that are highly optimized for specific alignment types, scoring schemes, and hardware targets with a single, uniform codebase (\autoref{overview} provides necessary background to understand the rest of the paper).
We show that using AnyDSL, we can design an alignment library that
\begin{itemize}
\item separates computation into two parts (a common part and an architecture-specific part) using higher-order abstractions without sacrificing performance (see \autoref{abstractions}),
\item provides mappings for CPUs using multithreading and vectorization without resorting to architecture-specific intrinsics, GPUs, and FPGAs using \ac{HLS} (see \autoref{mapping}),
\item allows for the variation of algorithmic parameters (e.g. type of alignment (global, local, or semi-global), scoring scheme (substitution matrix, linear or affine gap penalties), calculating only the optimal alignment score or an actual optimal alignment) by simple function composition rather than hard-to-understand metaprogramming techniques,
\item is competitive in terms of performance compared to state-of-the-art manually optimized sequence alignment libraries; that is, AnySeq is at most 7\% slower and in many cases (up to 12\%) faster than SeqAn on CPUs and NVBio on GPUs (see \autoref{performance}). In addition, we demonstrate that our proof-of-concept FPGA implementation runs more than 3 times (4 times) more energy efficient than the corresponding CPU (GPU) implementation.
\end{itemize}

\section{Overview}\label{overview}


\subsection{Motivation}
Parallelization of pairwise sequence alignment has been proposed on various architectures ranging from CPUs to GPUs and FPGAs. However, most implementations are tied to certain alignment scenarios, specific hardware, and execution models. This motivates the design of a flexible yet efficient sequence alignment library that abstracts a generic \ac{DP} algorithm and specializes only hardware-specific parts for a certain architecture.

To achieve high performance it is often inevitable to instantiate the code with different algorithmic variants and parameters and tailor it towards the target hardware architecture. Ideally, changing performance-impacting parameters should require minimum effort and have a negligible impact on size and complexity of the codebase. For example, the two distinct cases of constructing an actual alignment or just computing the optimal score, should not require two independent implementations of the alignment algorithm. Parts that are the same for both cases should use the same code and avoid duplication.

Existing alignment libraries (such as NVBio~\cite{nvbio} for GPUs, and Parasail~\cite{parasail}, SSW~\cite{ssw}, and SeqAn~\cite{reinert2017seqan} for CPUs) are optimized for a specific architecture. However, while especially GPU implementations differ greatly from their CPU counterparts due to the fundamentally different processor architectures, many parts of the algorithm are in principle the same for both platforms. When targeting different hardware architectures more than one programming language is necessary, e.\,g., \CC for the CPU code, CUDA for GPUs, or Verilog for FPGAs. This in turn leads to code duplication and increased code complexity.

We meet the challenges laid out above with the AnyDSL framework.
It provides the programming language Impala, which features a partial evaluator.
This approach allows for the design of the library AnySeq with high-level \emph{abstractions} that are successively instantiated with parameters and hardware-dependent code parts using partial evaluation instead of providing hand-tailored implementations for specific hardware and parameter sets as opposed to existing libraries.
This way, most code is generic and reusable across different hardware architectures and application scenarios to a substantial amount.

For example, DP matrices are not directly accessed as a 2D~array in AnySeq but through accessor functions that act as a \emph{view} on a 2D~data space. This decouples the implementation of the core algorithm from the concrete choice of the data layout.
In conventional settings, programmers would most likely refrain from such an implementation because it puts them into the hands of the compiler to succinctly optimize the runtime overhead of this indirection away.
In AnySeq we reliably direct the partial evaluator to remove any such overhead at compile time.

\subsection{AnyDSL Compiler Framework}
\label{sec:anydsl}

AnySeq is written in \impala, an imperative and functional language, which is part of the AnyDSL framework~\cite{DBLP:conf/gpce/LeissaBHMS15,DBLP:journals/pacmpl/LeissaBHPMSMS18}.

\paragraph{Partial Evaluation}
\label{sec:anydsl:pe}

\impala integrates a partial evaluator~\cite{DBLP:conf/rims/Futamura82}.
Programmers control the partial evaluator via \emph{filters}~\cite{DBLP:conf/esop/Consel88}.
These are Boolean expressions of the form \lst|@(expr)| that annotate function signatures.
\emph{Each} call site instantiates the callee's filter with the corresponding argument list.
If the expression evaluates to \texttt{true}, the call will be specialized.
Additionally, the expression \lst|?expr| yields \texttt{true}, if \lst|expr| is known at compile time;
the expression \lst|$\texttt\textdollar$expr| is never considered constant by the evaluator.
For example, the following
\lst$@(?n)$ filter will only
specialize calls to \lst[language=impala]$pow$
if \lst[language=impala]$n$ is statically known at compile time:
\begin{lstlisting}[language=impala]
fn @(?n) pow(x: int, n: int) -> int { /*...*/ }
\end{lstlisting}
Thus, the call \lst|pow(x, 5)| produces a loop-less sequence of multiplications,
\lst|pow(3, 5)| evaluates to \lst|243| while
\lst|pow(x, $\texttt\textdollar$5)| remains untouched.
This is called \emph{polyvariance}.
As syntactic sugar, \lst|@| is available as shorthand for \texttt{@(true)}.
This causes the partial evaluator to always specialize the annotated function.
What is more, \impala automatically annotates higher-order parameters for specialization.

\paragraph{Generators}

Because iteration is a common pattern, \impala provides syntactic sugar for invoking certain higher-order functions.
The loop
\begin{lstlisting}[language=impala]
for var1, ..., varn in iter(arg1, ..., argn) { /*...*/ }
\end{lstlisting}
translates to
\begin{lstlisting}[language=impala]
iter(arg1, ..., argn, |var1, ..., varn| { /*...*/ });
\end{lstlisting}
The body of the \texttt{for}-loop and the iteration variables constitute an anonymous function
\lst$|var1, ..., varn| { /*...*/ }$
that is passed to \lst|iter| as the last argument.
We call functions that are invokable like this \emph{generators}.
In particular, programmers can define their own iteration functions (see below).

\paragraph{Intrinsic Generators}
\label{sec:anydsl:intrinsics}

\impala exposes several forms of parallelism via built-in generators.
These do not possess an implementation in \impala itself but are recognized by the compiler.
The intrinsics used in our implementation are \lst|parallel| for spawning threads,
\lst|vectorize| for SIMD vectorization on the CPU,
\lst|cuda| for generating GPU code,
and \lst|hls| for \ac{HLS}.
The passed function is executed in a \ac{SPMD} context.
The following example demonstrates how to parallelize the outer and vectorize the inner loop in a 2D loop nest:
\begin{lstlisting}
for y in parallel(/*num threads*/ 4, y_beg, y_end) {
    for x in vectorize(/*simd width*/ 8, x_beg, x_end) {
        body(x, y)
    }
}
\end{lstlisting}

Most notably, they are not pragmas but regular functions that can be captured and passed around like any other function (see below).
The only restriction is that after partial evaluation, their higher-order arguments have to be function literals.


\paragraph{Custom Generators \& Partial Evaluation}

Generators are particularly powerful in combination with partial evaluation.
\begin{lstlisting}[mathescape=false]
type Body = fn(int) -> ();

fn @(?a & ?b) unroll(a: int, b: int, body: Body) -> () {
  if a < b {
    body(a);
    unroll(a+1, b, body)
  }
}

fn @range(a: int, b: int, body: Body) -> () {
  unroll($a, b, body)
}
\end{lstlisting}
Both functions are generators and, thus, are amenable to the \texttt{for}-syntax:
\begin{lstlisting}[language=impala]
for y in range(y_beg, y_end) {
    for x in unroll(0, 4) { body(x, y) }
}
\end{lstlisting}
They iterate from \lst|a| (inclusive) to \lst|b| (exclusive) while
invoking \lst|body| each time.
The filter of \lst|unroll| tells the partial evaluator to completely unroll the recursion if both loop bounds are statically known at a particular call site (as in the example above).
The function \lst|range| wraps a call to \lst|unroll| but prevents the partial evaluator from unrolling because it always considers \lst|$\texttt\textdollar$a| as dynamic.

Since generators are ordinary functions, programmers can pass them around, or combine them to build more sophisticated patterns.
The following example is parametric in an arbitrary 2D generator \mbox{\lst|loop2d|}.
\begin{lstlisting}
fn compute(/*...*/, loop2d: Loop2D) -> () {
  for x, y in loop2d(/*...*/) {  body(x, y) }
}
\end{lstlisting}
The function \mbox{\lst|combine|} creates 2D generators by composing two 1D generators.
This lets us explore various 2D loop nests that are feedable to \mbox{\lst|compute|}:
\begin{lstlisting}
let a = combine(range, range);
let b = combine(range, unroll);
let c = combine(|a, b, body| vectorize(8, a, b, body),
                |a, b, body| parallel (4, a, b, body));
let d = tile(64, 32, |a, b, body| vectorize(8, a, b, body),
                     |a, b, body| parellel (4, a, b, body));
\end{lstlisting}
The function \mbox{\lst|tile|} is a more sophisticated variant of \mbox{\lst|combine|} that sets up a tiled 2D loop nest.
Note that both \mbox{\lst|combine|} and \mbox{\lst|tile|} are ordinary functions that are implemented as library functions.
Partial evaluation will completely remove all overhead and generate a program that only consists of the desired loop nest while the function \mbox{\lst|compute|} is not tainted with any hardware-specific details.
Programming idioms like this are typical in our code base and showcase what is possible with AnyDSL's partial evaluator.




\section{Abstractions for Sequence Alignment}\label{abstractions}

AnySeq features a modular design that allows changing performance-impacting algorithm parameters at compile time:

\begin{itemize}
    \item hardware platform: CPU, CPU-SIMD, GPU, or FPGA
    \item alignment reconstruction needed: yes/no
    \item substitution function
    \item gap penalty scheme (linear or affine) and values
\end{itemize}

AnySeq achieves this via function composition, mostly in the form of providing behavior-controlling functions as arguments to higher-order functions (see \autoref{access_abstraction}).

\subsection{Pairwise Sequence Alignment}\label{alignment}


Consider two sequences $Q=(q_1 q_2 \ldots q_n)$ and $S=(s_1 s_2 \ldots s_m)$ of length $n$ and $m$ over the alphabet $\Sigma$. Their optimal alignment can be found in $\mathcal{O}(n \cdot m)$ by recursively solving three smaller subproblems. For each pair $(q_i,s_j)$ of characters one has to decide if these characters should be aligned or if a gap should be inserted.
The optimal alignment score $H(i,j)$ for the prefixes $(q_1 \ldots q_i)$ and $(s_1 \ldots s_j)$ is given by the recurrence relation
\begin{equation} \label{eq:H}
    H(i,j) = \max
    \begin{cases}
        H(i-1,j-1) + \sigma(q_i,s_j) \\
        E(i,j) \\
        F(i,j) \\
        \nu \\
    \end{cases}
\end{equation}
where $\sigma$ is a substitution function over $\Sigma \times \Sigma$ that determines the score of aligning two characters. The parameter $\nu$ is needed to distinguish between local and global alignments as we will explain below.

In case of a linear gap penalty $g$ we set:
\begin{align} \label{eq:lingap}
    E(i,j) &= H(i-1,j) - g \\
    F(i,j) &= H(i,j-1) - g
\end{align}

For affine gap penalties, a gap of length $k$ is penalized with $G_o + k \cdot G_e$ where $G_o$ is the cost of opening a gap and $G_e$ is the cost of extending a gap. We then need two additional auxiliary \ac{DP} matrices to determine the optimal value for $E$ and $F$ at position $(i,j)$. The following functions determine the best alignment score for the prefixes $(q_1 \ldots q_i)$  and $(s_1 \ldots s_j)$ under the constraint that $s_j$ (or $q_i$) is aligned to a gap:
\begin{align}
\label{eq:E}
    E(i,j) &= \max
    \begin{cases}
        E(i-1,j) - G_e \\
        H(i-1,j) - G_o - G_e
    \end{cases} \\
\label{eq:F}
    F(i,j) &= \max
    \begin{cases}
        F(i,j-1) - G_e \\
        H(i,j-1) - G_o - G_e.
    \end{cases}
\end{align}

Initialization of the first rows and columns of $H$, $E$, and $F$---as well as in what cell(s) to look for the optimal score---depends on whether the alignment shall be \emph{global}, \emph{local}, or \emph{semi-global}. Note that for the following equations holds $1 \leq i \leq n, 1 \leq j \leq m$; the initialized rows and columns have index $0$.

In the case of computing an optimal {\em local alignment}, we look for the best scoring alignment starting at any position $(q_i,s_j)$ and ending at any other position $(s_k,q_l)$ with $i \leq k \leq n, j \leq l \leq m$. The parameter $\nu$ in \autoref{eq:H} has to be set to~$0$ in order to ensure that scores will always be positive. The \ac{DP} matrices are initialized as follows: $H(0,0) = H(i,0) = H(0,j) = 0$, $E(0,0) = F(0,0) = F(i,0) = E(0,j) = -\infty$, $E(i,0) = -G_o -i \cdot G_e$, and $F(0,j) = -G_o - j \cdot G_e$.


{\em Global alignments} always start at position~$(0,0)$ and end at position~$(n,m)$. Hence, the optimal score lies in cell $H(n,m)$. In order to model unbounded scores, the parameter $\nu$ in \autoref{eq:H} must be~$-\infty$. The \ac{DP} matrices are initialized as follows: $H(0,0) = 0$, $E(0,0) = F(0,0) = F(i,0) = E(0,j) = -\infty$, $H(i,0) = E(i,0) = -G_o -i \cdot G_e$, and $H(0,j) = F(0,j) = -G_o - j \cdot G_e$.

For {\em semi-global alignments}, gaps at the beginning and at the end are not penalized. This leads to the same initialization as for local alignments. The optimal score lies in the last row or column of $H$.

Note that score-only computations can be performed in linear space and quadratic time with respect to the length of the alignment targets. Actual alignments producing this value can be constructed by \emph{tracing back} the predecessor information in the \ac{DP} matrices. In order to avoid quadratic space consumption (which is prohibitive for long DNA strings), the \emph{traceback} procedure can be implemented in linear space by a divide-and-conquer approach~\cite{hirschberg75} that recursively determines optimal midpoints of the \ac{DP} matrix (at the cost of at most doubling the amount of computed \ac{DP} cells).

\begin{figure}[t]
    \centerline{\includegraphics[width=0.46\textwidth]{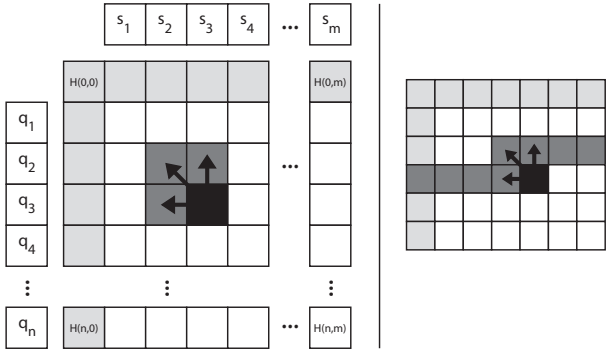}}
    \caption{The left image shows the conceptual \ac{DP} matrix $H$. Light gray cells are initialization cells and dark gray indicates the ancestral subproblems of the currently active cell, which is shown in black. The right image shows the cells that need to be stored for score-only computation in dark gray.}
    \label{fig:dpmatrix}
\end{figure}

Each cell of $H$ in \autoref{eq:H} depends on three neighboring cells (see \autoref{fig:dpmatrix}) which means that relaxing them in parallel can be done along minor diagonals.
When relaxing all cells in a submatrix row of $H$, only the subproblem scores for the row above and one column left of the current cell are needed. If we want to compute submatrices in parallel, the first row and first column of a submatrix must have been computed earlier and the last row and last column must be kept available for the computation of subsequent submatrices (to the right of and below the current one).

\subsection{Data Access Abstractions}\label{access_abstraction}

Efficient scalar CPU implementations need radically different iteration and blocking schemes than efficient SIMD-vectorized CPU implementations or efficient GPU code. This in turn leads to different data access patterns and data storage requirements (memory alignment, RAM vs. VideoRAM, GPU shared memory, etc.). A key idea of our design is the usage of data accessor objects for decoupling data access from the actual data storage strategy.

For example, while the resulting data indexing schemes may differ, the underlying relaxation equations \eqref{eq:H}, \eqref{eq:E}, \eqref{eq:F} stay the same. It is therefore desirable to decouple these two aspects of the algorithm. AnySeq encodes the basic recurrence relation for one \ac{DP} matrix cell update in one function. The following example showcases this for global alignments:
\begin{lstlisting}
fn relax_global(
    scoring: Scoring,  // scoring scheme, may access E and F
    prev: PrevScores,  // accessor to previous  scores
    next: CharPair     // accessor to sequence letter pair
    ) -> NextStep      // optimum score and predecessor
{
  let mut res = NextStep {              // no gaps
    score: scoring.subst(prev, next),
    predc: PRED_NO_GAP
  };

  let sgap = scoring.gap_s(prev, next); // subject gap
  if sgap > res.score {
    res.score = sgap;
    res.predc = PRED_SKIP_S;
  }

  let qgap = scoring.gap_q(prev, next); // query gap
  if qgap > res.score {
    res.score = qgap;
    res.predc = PRED_SKIP_Q;
  }

  res // return maximum over all three choices
}
\end{lstlisting}

Member functions of type \lst|PrevScores| are used to access scoring information of the three ancestral subproblems. If no gap should be introduced, the scoring substitution function is used to determine the cost. In case of affine gap penalties the functions \lst|gap_s| and \lst|gap_q| will access the auxiliary matrix elements from $E$ and $F$, for linear penalties they return a constant. Objects of type \lst|NextStep| store the maximum score for the current cell and what previous subproblem was chosen as ancestor. The ancestor information encoded in \lst|res.predc| is sometimes needed for the innermost level of the traceback since recursion on subsequences is only done if the subsequence sizes exceed a hardware-specific threshold. Whether any of the functions access main memory, GPU memory, or simply returns a constant is determined based on the algorithm parameterization at compile time. The relaxation order (which for the recursive traceback is reversed for half of the \ac{DP}  matrix) and subproblem sizes are also determined based on algorithmic parameters and the target hardware platform.

Accesses to the values of previously computed subproblems as well as to the input sequence characters are also abstracted by function calls. This makes it possible to vary intermediate result storage strategies, indexing, and blocking schemes independently of other parts of the algorithm:
\begin{lstlisting}
// read-only access to a sequence of characters
struct Sequence {
  len:     fn() -> Index,
  at:      fn(Index) -> Char,
  release: fn() -> ()
}
// read/write access to a sequence of characters
struct MutableSequence {
  len:     fn() -> Index,
  at:      fn(Index) -> Char,
  write:   fn(Index, Char) -> (),
  release: fn() -> ()
}
// access to alignment scores
struct Scores {
  prev:    fn() -> PrevScores,
  at:      fn(IndexPair) -> Score,
  update:  fn(IndexPair, Score) -> (),
  release: fn() -> ()
}
\end{lstlisting}

AnySeq resorts to generators to define indexing/blocking schemes.
These are used in CPU as well as in GPU code.
This is explained in more detail in the following subsections.

Since AnyDSL supports \textit{partial evaluation} and the function call hierarchy is known at compile time, AnySeq can use several layers of indirection without impacting performance. This allows us to isolate hardware-specific parts from architecture-independent code such as the relaxation function in a way that is much easier to write, understand, and reason about than for \CC~templates.

\subsection{Interface \& Data Flow}\label{interface}

The outermost interface functions build accessor objects to the sequence storage and combine different initialization, relaxation, and matrix traversal/blocking functions to realize different alignment schemes. To make interfacing with other languages easier, AnySeq provides C wrapper functions as entry points to the different algorithmic parameterization scenarios:
\begin{lstlisting}
extern fn construct_global_alignment(
        query: &[u8], subj: &[u8], // sequence input
        qAlign: &mut [u8],         // alignment output
        sAlign: &mut [u8]
    ) -> Score                     // optimum score
{  // yields data accessor objects to input and output
  let input  = input_view(query, subj);
  let output = output_view(qAlign, sAlign);

  // yields a struct with functions to control algo behavior
  let scheme = global_scheme(
        linear_gap_scoring(simple_subst_scoring(2,-1), -1));
  construct_alignment(scheme, input, output)
}
\end{lstlisting}
In order to use it in \CC, it just needs to be declared:
\begin{lstlisting}[language=C]
extern "C" { // C linkage
score_t construct_global_alignment(const char* query,
  const char* subj, char* qAlign, char* sAlign);
}
\end{lstlisting}

Algorithmic variants can be obtained by exchanging parameters of higher-order functions. Scoring schemes, memory access, and iteration strategies are all encapsulated in functions that can be interchanged. Most of the time programmers need to exchange several functions in order to get the desired behavior. Thus, functions are often grouped into descriptively named structs to make parameterization more convenient and the resulting code more expressive.

Because all accesses to actual data are abstracted through function calls we can simply exchange the order in which memory is accessed. If, for example, parts of the input sequences need to be reversed (this is the case for the divide-and-conquer traceback) we reverse the indexing in the sequence accessor function.

The following code example shows accessors that decouple memory access patterns. A \lst|MatrixView| instance provides an interface for reading and writing data addressed by two indices. Functions like \lst|view_matrix_coal_offset| create accessors that are optimized for a specific hardware architecture or algorithmic stage.
\begin{lstlisting}
type MatrixReadFn  = fn(Index, Index) -> Score32;
type MatrixWriteFn = fn(Index, Index, Score32);

struct MatrixView {
  read:  MatrixReadFn,
  write: MatrixWriteFn
}
// coalescled access pattern for GPUs
fn view_matrix_coal_offset(matrix: Matrix,
    read: MatrixReadFn, write: MatrixWriteFn,
    oi: Index, oj: Index) -> MatrixView
{
  let coalesced_pos = |i: Index, j: Index| -> Index {
    ((i + oi + j + oj + 2) % matrix.mem_height)
     * matrix.mem_width + j + oj
  };
  MatrixView {
    read:  |i, j| read(coalesced_pos(i, j)),
    write: |i, j, value| write(coalesced_pos(i, j), value)
  }
}
\end{lstlisting}

Scoring schemes are built by combining a gap scoring strategy with a substitution function (see the listing in \autoref{interface}). The substitution function takes an accessor to the scores of the ancestral problems and the next character pair.
We obtain a simple scoring function with a match score of~$2$ and a mismatch score of~$-1$ by calling
\lst|simple_subst_scoring(2, -1)|.
It returns the corresponding substitution function.
\begin{lstlisting}
// signature of a scoring function
type ScoringFn = fn(PreviousScores, CharPair) -> Score;

// make scoring function for a simple scheme
fn simple_subst_scoring(same: Score, mismatch: Score)
-> ScoringFn {
  // this lambda expression returns a function
  |prev, chars| {
    prev.no_gap +
      if chars.q == chars.s { same } else { mismatch }
  }
}
\end{lstlisting}
This example also demonstrates how the ability to return functions makes parameterizations easier to use at the call site.
To obtain a matrix substitution scheme, the programmer simply has to pass a substitution function that reads scores from a lookup table.
AnySeq implements gap scoring functions similarly.

Since partial evaluation ensures that no machine code is generated for calls to functions that either do not contain instructions or return a compile-time constant, we can implement most algorithmic details in their most general form and rely on partial evaluation to eliminate such calls. For example, if the predecessor information or affine gap storage is not needed in a concrete algorithmic variant, the functions in the specialized accessor objects do either nothing or return compile-time constant values. The same holds for writing score data. For local and semi-global alignment computation the algorithm must also keep track of the current maximum score while this is not necessary for global alignments. For this, we just have to swap out one variant of the \lst|Scores| accessor's member function \lst|update| (see \autoref{access_abstraction}) for a different (more efficient) one at compile time.


The resulting control flow for computing an alignment is as follows:
\begin{enumerate*}
    \item allocate memory for the input data,
    \item read and store input data,
    \item allocate storage needed for the output,
    \item build accessors to input and output data storage,
    \item allocate temporary storage for the alignment algorithm (RAM, GPU global and/or shared memory),
    \item build accessors to the temporary storage(s),
    \item run relaxation procedure for computing the \ac{DP} matrix cells,
    \item look up optimal score,
    \item if needed, build alignment strings,
    \item output results
\end{enumerate*}.

\section{Mappings to CPUs, GPUs, and FPGAs}\label{mapping}

If high performance is desired, it becomes inevitable that some parts of an algorithm need to be specialized for different acceleration technologies.
One way of supporting different hardware platforms, is to choose one of several implementations of the same struct or function, i.\,e., by writing different, specialized variants with the same name or signature, respectively. As an example, consider the following CPU version of a function \lst|inplace_map| that transforms the array \lst|data| within the range \lst|a| to \lst|b| by applying a function \lst|f|:
\begin{lstlisting}
fn inplace_map(a: int, b: int, data: &mut [float],
               f: fn(float) -> float) -> () {
  for i in range(a, b) {
    data(i) = f(data(i)); // apply f to array value i
  }
}
\end{lstlisting}
A GPU implementation (that does not use shared memory) of the same function looks like this:
\begin{lstlisting}
fn inplace_map(/*as above*/) -> () {
  // determine optimum block and grid dimensions
  let block_width = opt_block_width(a, b);
  let block = (block_width, 1, 1);
  let grid  = (opt_num_blocks(a, b) * block_width, 1, 1);
  with cuda(grid, block) { // run kernel
    let i = cuda_gidx();   // get global id
    data(i) = f(data(i));  // apply transformation
  }
}
\end{lstlisting}
The corresponding FPGA implementation uses a pipelined execution:
\begin{lstlisting}
fn inplace_map(/*as above*/) -> () {
  with hls() {
    for i in pipelined(a, b) {
      data(i) = f(data(i));
    }
  }
}
\end{lstlisting}

The interface of \lst|inplace_map| can be made more generic by using data accessor functions instead of pointers to arrays.

We have used hardware-specific generator functions to encapsulate iteration and blocking strategies
and data accessors to encapsulate memory access patterns. Therefore, supporting a new hardware platform means replacing only those iteration and data access abstractions for which benchmarks have shown that they perform sub-optimally on that platform.
A breakdown of our code base---excluding supporting code like benchmarking functions, I/O, and C interfacing functions and also the FPGA-specific parts---reveals that approximately 23\% of all lines of code are specifically written for the GPU, 14\% are specific to CPU vectorization and less than 11\% are only needed for the non-vectorized CPU version while the remaining 52\% are shared among all three variants.




\subsection{CPU Parallelization}

We compute the values of different \ac{DP} submatrices concurrently using CPU threads. As mentioned before, relaxing \ac{DP} matrix cells in parallel can be done along (minor) diagonals. In the non-vectorized version, cells within a submatrix will be relaxed in row-major order.

\begin{figure}[t]
    \centerline{\includegraphics[width=0.1\textwidth]{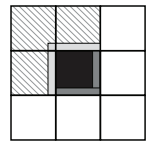}}
    \caption{Data dependencies for submatrix relaxation. Here hatched submatrices have already been computed. Border cells marked light gray need to be communicated to the currently active, black submatrix. Dark gray cells need to be stored for the submatrices that depend on the current one. }
    \label{fig:submatrix}
\end{figure}

Zero-overhead data access abstractions allow for the convenient separation of iteration logic and memory layout. Note that scores must be accessed following a row-major order within a tile. This is because if we want to compute the value at \emph{local} position $(i,j)$ an intra-tile cyclic buffer must always contain the previously computed values at \emph{local} positions $(i-1,j-1)$ through $(i-1,m)$ and $(i,0)$ through $(i,j-1)$ (see the right image in \autoref{fig:dpmatrix}). Furthermore, the values of the rightmost and bottommost border cells of a submatrix need to be kept as long as neighboring submatrices that depend on these values have not been computed yet (see \autoref{fig:submatrix}). Again, data accessors help to hide the fact that not the entire \ac{DP} matrix is stored, but only such border stripes.

In our preliminary version of AnySeq \cite{DBLP:journals/pacmpl/LeissaBHPMSMS18} we used a static wavefront schedule along diagonals of submatrices and vectorization was done by computing consecutive rows within a submatrix while substitution scores were looked up from a precomputed auxiliary array. This approach did not yield satisfactory performance (see red line in \autoref{fig:scaling}).

Instead of a static schedule we now use a dynamic wavefront approach where submatrices are scheduled in a thread-safe queue which allows threads to add and extract work items concurrently. Not only does this eliminate load imbalances due to a mismatch between the number of available threads and the number of submatrices that can be relaxed in parallel, this approach also leads to better load-balancing when computing several alignments of different sizes concurrently.

Vectorization is done over blocks that consist of rows from independent submatrices which also obviates the need for an auxiliary data structure for score lookup. The number of rows in each block corresponds to the number of available vector lanes $l$ which is determined by the SIMD instruction set and the width of the data type used for computing score values.

\impala provides the generator \lst|vectorize| that triggers CPU vector code generation which in turn relies on the Region Vectorizer~\cite{DBLP:conf/pldi/MollH18}. Within the body of \lst|vectorize| memory reads and writes as well as arithmetic operations are transformed into the corresponding vector instructions. Most existing vectorized alignment implementations~\cite{swaphils, parasail, ssw, misra2018performance, hou2016aalign} are based on intrinsics specific to certain architectures. A major advantage of our approach is that the \lst|vectorize| generator supports several SIMD instruction sets.

Since only differences to the global score are relevant, we use smaller data types (e.g.\ 16~bits for our use cases) for scores within a block. Whether this is feasible without over- or underflow, depends on the block size and the scoring scheme. The largest possible differential score value occurs if all characters in the corresponding subalignment match. The smallest possible differential score is obtained if no pair of characters in both sequences matches and either the largest possible mismatch penalty (along the alignment diagonal) or the largest possible gap penalty (along the first row or first column) is subtracted.

\begin{figure}[t]
    \centerline{\includegraphics[width=0.4\textwidth]{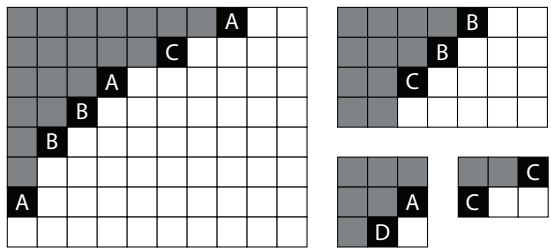}}
    \caption{
        Submatrix scheduling example: four alignments of different sizes are computed
        simultanously by distributing independent subalignments over several
        threads. Subalignments that are already finished are shown in grey,
        currently active ones are shown in black. Each of threads A, B, and C
        computes four independent subalignments using vectorization ($l$=4)
        while thread D computes one subalignment using the scalar method.
    }
    \label{fig:simd}
\end{figure}

A thread only computes a vectorized block, if $l$ work items are enqueued. This means that the preconditions of $l$ independent alignment submatrices already hold. Depending on the workload, e.g., when computing alignments of several sequence pairs in parallel, or when starting new alignments, there might be less than $l$ submatrices available per thread. In these cases threads will compute single submatrices using the scalar method (see  \autoref{fig:simd}). After computation of a block has finished, all of the $l$ associated alignment submatrices are marked as finished by the current thread. Next, for each of the completed submatrices, their neighboring submatrices are enqueued if they have neither been computed nor been enqueued yet and their preconditions have already been fulfilled, i.\,e., their predecessor submatrices have been computed. The completion and queuing status of all submatrices is tracked using preallocated arrays of atomic flags.

\subsection{GPU Parallelization}

\begin{figure}[t]
    \centerline{\includegraphics[width=0.25\textwidth]{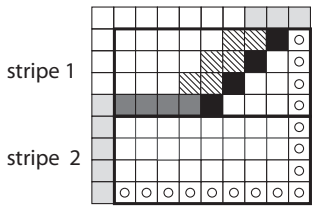}}
    \caption{Striped computation of a tile on the GPU.
    Values of the cells shown in gray and hatched cells are located in shared memory.
    Light gray indicates values still needed for the current tile. Dark gray indicates values needed for the next stripe (re-use memory previously claimed by initialization cells that are no longer needed). The circles indicate values that need to be written to global memory after the current tile has been completed.}
    \label{fig:gpu}
\end{figure}

Similar to the CPU code we iterate over matrix tiles in a diagonal fashion. This is done in host code that starts a GPU kernel for each diagonal. The GPU kernel uses a one-dimensional grid of thread-blocks where each block computes one matrix tile. Each tile is further divided into stripes which are computed in sequence by a thread-block. Within a stripe, threads compute diagonals in parallel (see \autoref{fig:gpu}). Computation of one stripe is divided into three parts in order to avoid branch divergence. This division is due to the fact that full diagonals are only computed in the middle part of a stripe but not at the start or the end.

Again, data accessors are used to manage memory buffers that can now point to global or block-local shared GPU memory. In order to enable coalesced memory access we have to use memory access patterns different from the CPU code. Since memory access is abstracted by functions, it is easy to exchange access schemes for the CPU and the GPU code. Device-independent code, like the core relaxation function, remain the same, because it only calls those memory access functions to read and write values.

Before starting a block of threads, the segments of the input sequences that are needed for the current tile are stored in block-local shared memory. The values of the row immediately above the current stripe are needed to access ancestral problems. Also the values in the bottom-most row of the current stripe will be needed by the next stripe. Since we are relaxing in diagonals along the stripe, we re-use the memory cells with the values of the uppermost row that are no longer needed and store the results of the last row to them. The buffer needed for this does also reside in shared memory. This way, we compute all stripes within a tile in succession without starting a new kernel. The last row and column of the current tile are always written back to global memory (indicated by circles in \autoref{fig:gpu}).

\subsection{FPGA Parallelization}

The FPGA implementation iterates over matrix tiles in a diagonal fashion, too.
We use several processing elements for this that all compute a single \ac{DP} cell per clock cycle.
The iteration in diagonal fashion happens in stripes of width $K_{\mathtt{PE}}$, i.\,e., the number of processing elements.
In turn, for each clock cycle we have $K_{\mathtt{PE}}$ updates in parallel, which renders $K_{\mathtt{PE}}$ and the frequency crucial for performance.
The shorter sequence is divided into blocks of maximum size $K_{\mathtt{PE}}$ which are used to initialize the processing elements.
Characters of the longer sequence are streamed one-by-one through the linear array of processing elements; that is, processing elements perform the relaxation and pass the character and their results to the next processing element.
The last element emits the score to the host system.

In case the shorter sequence is longer than $K_{\mathtt{PE}}$, we buffer the rightmost \ac{DP} column of a stripe with the help of a predefined hardware component in DDR memory of the host system.
This component is necessary to store the results in parallel to the ongoing computation. We have similar components to stream the characters and for fetching the previously stored values from DDR.

\section{Performance Evaluation}\label{performance}

Experiments were run on a system with two Intel Xeon Gold 6130 (Skylake) CPUs with 192\,GB of DDR4 RAM and L1, L2, and L3 caches sizes of 1\,MB, 16\,MB, and 22\,MB respectively.
Each CPU has 16 physical cores and all CPU algorithms were run with 32~threads.
The operating system used was CentOS Linux release 7.6.1810.
The vectorized variants were compiled for the AVX2 and AVX512 instruction sets and use 16~bit scores within a SIMD lane.
The GPU experiments were performed on a Titan V.

We compared AnySeq to the well-established sequence alignment libraries SeqAn~2.4.0, Parasail~2.0, and NVBio~1.1.
AnySeq was compiled with the AnyDSL version from July 2019 and libraries were compiled with gcc~7.3.0.
AnyDSL is based on LLVM~8 and links against CUDA~10.1.
After some modifications, NVBio could be compiled with CUDA~8.

We have evaluated performance for two common use cases: (i)  pairwise  alignment  for  long  DNA sequences and (ii) comparisons of large amounts of short Illumina reads. We mostly computed global alignments using a simple scoring scheme with $+2$ for a match, $-1$ for a mismatch and a linear gap penalty of $-1$ and in addition global alignments with an affine gap scoring scheme using $G_o=-2$ and $G_e=-1$ if supported. Parasail does not explicitly specialize the case of linear gap penalties which means that it effectively always computes affine gaps, even if $G_o=0$.

\begin{table}[t]
\centering
\resizebox{\columnwidth}{!}{%
\begin{tabular}{lrl}
\toprule
  Accession No. &  Length     &  Genome Definition                 \\
\midrule
  NC\_000962.3        & 4,411,532   &  Mycobacterium tuberculosis H37Rv  \\
  NC\_000913.3        & 4,641,652   &  Escherichia coli K12 MG1655       \\
  NT\_033779.4        & 23,011,544  &  Drosophila melanogaster chr. 2L   \\
  BA000046.3          & 32,799,110  &  Pan troglodytes DNA chr. 22       \\
  NC\_019481.1        & 42,034,648  &  Ovis aries breed Texel chr. 24    \\
  NC\_019478.1        & 50,073,674  &  Ovis aries breed Texel chr. 21    \\
\bottomrule
\end{tabular}
}%
\caption{Long genomic sequences used for benchmarking}
\label{tab:genomes}
\end{table}

\begin{figure*}[t]
    \captionsetup{margin={3cm,0cm}}
    \input{bars.tex}
    \label{fig:performance}
\end{figure*}

For the first use case we aligned three pairs of long genomic sequences of roughly similar length listed in \autoref{tab:genomes}, which have been used as a benchmark before \cite{swaphils, rahn2018generic}. Part a) of \autoref{fig:performance} shows the median performance in terms of \ac{GCUPS} on the tested CPU and GPUs for both score-only computation and traceback. For score-only computation AnySeq is faster than both SeqAn and Parasail for the multithreaded, non-vectorized version, and for the AVX2 version while for the AVX512 version SeqAn is faster. For the traceback, AnySeq and SeqAn have roughly the same speed for all versions---both outperforming Parasail. On the tested GPU, AnySeq outperforms NVBio for both score-only computation and alignment reconstruction by a factor of up to 1.1. Using affine gap penalties does generally not change the relative performances of the tested libraries, although it is always slower than using linear gap penalties due to an increased amount of memory reads and writes. Note that alignment computation on the GPU relies on 32-bit integer arithmetic due to the lack of native support for shorter integer data types (as used by our AVX vectorization).

For the second use case we performed 12.5~million pairwise alignments of Illumina reads of length 150 base-pairs each. The set of reads was simulated with Mason using chromosome 10 of GRCH38 as reference. Part b) of \autoref{fig:performance} shows the achieved performance in \ac{GCUPS} on the tested CPU and GPU for both score-only computation and traceback. For score-only computation on the CPU, AnySeq outperforms both SeqAn and Parasail for the AVX2 version, while for the AVX512 version SeqAn is slightly faster. On the GPU, AnySeq consistently outperforms NVBio with a factor of up to~1.12.

\begin{figure}[t]
    \centering
    \begin{tikzpicture}
        \tikzstyle{every node}=[font=\small]
        \begin{axis}[
                width=0.75\columnwidth,
                xmode=log,log basis x={2},
                ymode=log,log basis y={10},
                log ticks with fixed point,
                xlabel={\# threads}, xlabel style={yshift= 1ex},
                ylabel={GCUPS},      ylabel style={yshift=-4ex},
                legend pos=north west,
            ]
            \addplot coordinates {  (1, 4.5) (2, 5.6) (4, 6.8) (8, 8.9) (16, 10.7) (32, 11.8) };
            \addlegendentry{static wavefront}
            \addplot coordinates { (1, 4.5) (2, 8.9) (4, 16.5) (8, 29.9) (16, 54.2) (32, 93.1) };
            \addlegendentry{dynamic wafefront}
        \end{axis}
    \end{tikzpicture}
    \caption{Thread scalability comparison on a CPU between dynamic wavefront and static wavefront parallelization for aligning a pair of long DNA strings using AVX2.}
    \label{fig:scaling}
\end{figure}

We have also compared the CPU thread scalability of the proposed dynamic wavefront approach with a purely static wavefront along diagonals in \autoref{fig:scaling}.
Our preliminary version~\cite{DBLP:journals/pacmpl/LeissaBHPMSMS18} and Parasail rely on the latter strategy.
This also explains the low Parasail performance in \autoref{fig:performance} part a).
The dynamic approach achieves an efficiency of 75\% and 65\% for 16 and 32 threads, respectively, while the corresponding efficiencies of the static approach are merely 15\% and 8\%.
SeqAn is also based upon a dynamic wavefront approach but relies on low-level intrinsics for vectorization to support various instructions sets.
This makes this code not only hard to port to different SIMD architectures but also requires to emulate control flow constructs such as \texttt{if}, \texttt{while}, or \texttt{break} with masked data flow---a time-consuming and error-prone process.

AnySeq slightly outperforms SeqAn and NVBio in several use cases due to different implementation details like the internals of the concurrent queue used for scheduling tiles or different parameter choices for recursion cutoff points or tile sizes.
That being said it is hard to pinpoint the exact reason that attributes to performance differences of about~5\%:
The code bases are very different and even slight differences in the generated code can easily impact performance.

\subsection*{FPGA Performance}

\begin{table}[t]
\centering
\begin{tabular}{lrlr}
    \toprule
    Device                                  & Watt                                     & Gap    & GCUPS/watt \\
    \midrule
    \multirow{2}{*}{Intel Xeon Gold 6130}   & \multirow{2}{*}{125\phantom{.000}$^\textnormal{a)}$}   & linear & 1.024 \\
                                            &                                          & affine & 0.968 \\
    \multirow{2}{*}{Titan V}                & \multirow{2}{*}{250\phantom{.000}$^\textnormal{a)}$}   & linear & 0.757 \\
                                            &                                          & affine & 0.696 \\
    \multirow{2}{*}{ZCU104}                 & \multirow{2}{*}{6.181$^\textnormal{b)}$} & linear & 3.187 \\
                                            &                                          & affine & 3.187 \\
    \bottomrule
\end{tabular}
\vspace{0.5ex}
\begin{footnotesize}%
\begin{flushleft}%
$^\textnormal{a)}$\,according to specification \\
$^\textnormal{b)}$\,according to hardware synthesis report 
\end{flushleft}%
\end{footnotesize}%
\vspace{-0.5ex}
\caption{
    Energy efficiency in terms of GCUPS/watt (higher is better) of AnySeq for all tested devices (scores only, long genomes).
    Baseline forms the fastest corresponding AnySeq variant from \autoref{fig:performance} for that device.
}%
\label{tab:fpga_power}
\end{table}

Our FPGA implementation is not as mature as the CPU or GPU one and---at the time of writing---only supports score-only long genome alignment.
We chose the Xilinx Zynq UltraScale+ MPSoC ZCU104 Evaluation Kit for our experiments.
AnySeq runs with a frequency of 187.5\,MHz
and achieves a median performance of about 20\,\ac{GCUPS} on the ZCU104 (see \autoref{fig:performance}; long genomes, scores only).

The runtime is not affected by the gap penalty scheme as the computation happens in a single clock-cycle nonetheless.
Moreover, without modifying the data transfer methodology, a no-operation hardware module is as fast as our alignment core.
This indicates that our implementation only improves if the data transfer rate increases.

Even though the total \ac{GCUPS} of the ZCU104 are not competitive to the tested high-end CPU/GPU systems, low-end FPGAs consume significantly less power. \autoref{tab:fpga_power} compares the power efficiency for each tested device in terms of achieved GCUPS \emph{per} watt.
The ZCU104 is more than 3 times more energy efficient than the corresponding CPU implementations and
4.2--4.5 times more energy efficient than the corresponding GPU implementations.

\section{Related Work}\label{relwork}

Parallelization of genomic sequence alignment has been investigated mainly in the context of two types of application  scenarios:  (i)  pairwise  alignment  for  long  DNA sequences, and (ii) comparison of large amounts NGS reads. Existing monolithic implementations and libraries have been optimized for specific architectures including CPUs, GPUs, and FPGAs. Two widely adopted parallelization schemes are \textit{inter-sequence} (computes \ac{DP} matrix cells of a number of independent alignment tasks in parallel) and \textit{intra-sequence} (computes \ac{DP} matrix cells of a single alignment in parallel).

FPGA solutions using systolic arrays \cite{oliver2005hyper}, early approaches using SIMD registers of standard CPUs \cite{wozniak1997using},  and  a number of GPU  approaches \cite{liu2013cudasw++} are based on the intra-sequence method vectorizing over minor diagonals of the \ac{DP} matrix. Farrar’s approach  \cite{farrar2006striped} is a popular  intra-sequence method  using a striped layout for  SIMD  registers. Unfortunately, its  performance  relies  on  efficient  branch  prediction units which are often inefficient on modern many-core architectures. Both multithreading and vectorization have been effectively employed on CPUs for the inter-sequence approach if a large number of alignments need to be computed in parallel \cite{misra2018performance}. This is typically the case when processing large-scale NGS datasets. However, for the alignment of a pair of long genomic sequences, most approaches employ a wavefront pattern for intra-sequence parallelization \cite{cudalign, korpar2013sw, rucci2018swifold, hou2016aalign, swaphils}. An additional level of coarse-grained parallelism (e.g. on CPU/GPU clusters) has been proposed based on speculative execution \cite{de2016cudalign, Maleki}.

These approaches have in common that they are based on a highly optimized architecture-specific kernel but are often not flexible enough to support different alignment scenarios and are not compatible to different architectures. Consequently,  libraries have been introduced that expose alignment algorithms as reusable components. Among them SeqAn \cite{reinert2017seqan} and NVBio \cite{nvbio} provide the highest flexibility on CPUs and GPUs, respectively. A recent paper \cite{rahn2018generic} has shown that the performance of SeqAn is superior to other CPU-based libraries including Parasail \cite{parasail} and SSW \cite{ssw}. Unfortunately, existing approaches are specific to certain architectures and execution models and are therefore not able to support a variety of modern hardware required for modern bioinformatics pipelines that need to deal with fast-growing biological sequence databases. AnySeq demonstrates that using partial evaluation it is possible to build an alignment library based on higher-order abstractions that can specialize on a variety of architectures (CPUs, GPUs, and FPGAs) with comparable performance to state-of-the-art libraries for two application scenarios (long sequence alignment and NGS read comparison) that are hand-optimized for a specific architecture (e.g. NVBio on GPUs, SeqAn on CPUs).

Both SeqAn and NVBio use \CC metaprogramming in order to specialize algorithmic variants at compile time.
Some projects even use scripts (e.g.\ written in Python or Ruby) that generate other programs (e.g.\ in C/\CC or CUDA).
However, metaprogramming entails severe drawbacks in terms of programmer productivity:
First, the meta language is different than the core language and/or intrusively pervades the program with quoted code snippets of the program to be generated (\CC template language vs.\ \CC core).
For this reason, metaprograms are hard to read, write, and understand.
What is more, programmers cannot easily move code between the core and the metaprogram.
Thus, they must manually implement all needed versions like \lst|pow(a, b)| vs.\ \lst|pow<b>(a)| vs.\ \lst|pow<a, b>()|.
In addition, C++ templates do not accept lambda expressions as arguments.
Finally, running the metaprogram does not guarantee the well-typedness of the generated program. 
This leads to the infamous, difficult to understand error messages \emph{after} \CC template instantiation.

AnyDSL on the other hand uses partial evaluation and does not suffer from any of these problems.
The only annotations required are filters that determine whether the evaluator should specialize a particular call site.
This allows in contrast to metaprogramming for polyvariance and avoids the need to implement several variants of the very same function (see \autoref{sec:anydsl}).
This paper displays in several examples that we pass functions \emph{with} free variables (lambda expressions in \CC terminology) around and later specialize these calls via partial evaluation.
This is not possible with \CC.
Finally, the partial evaluator runs in contrast to template instantiation on a well-typed program and, hence, will not produce an ill-typed one.
Please refer to Leißa et.\ al.~\cite[§3]{DBLP:journals/pacmpl/LeissaBHPMSMS18} for a thorough discussion of metaprogramming and partial evaluation.


\section{Conclusion}\label{conclusion}

The continually increasing volume of genomic sequencing data poses severe challenges when developing bioinformatics methods of practical relevance. To achieve sufficient performance frequently requires the use of modern hardware architectures. One way to separate the concerns of bioinformatics method development and low-level parallelization and optimization is the use of sequence alignment libraries. Unfortunately, existing approaches such as SeqAn, Parasail, or NVBio only provide optimized implementations for a single architecture and add significant code complexity through template metaprogramming.

We have presented a new approach for designing a high performance library for genomic sequence alignment based on partial evaluation. By using higher-order abstractions, \emph{AnySeq} separates the computation into common parts implemented in a generic way and parts allowing for architecture-specific optimization. Implementations of different alignment variants by simple function composition and mappings for CPU, GPU, and FPGA-based hardware targets have been presented. Through the use of recent compiler technology that incorporates partial evaluation any overhead incurred by the utilized higher level abstractions can be effectively eliminated. As a result AnySeq achieves highly competitive performance comparable to a number of state-of-the-art, hand-optimized alignment libraries on various platforms. AnySeq is open source software and can be downloaded at \url{https://github.com/AnyDSL/anyseq}.




\bibliographystyle{IEEEtran}
\bibliography{bibliography}

\begin{thebibliography}{10}
\providecommand{\url}[1]{#1}
\csname url@samestyle\endcsname
\providecommand{\newblock}{\relax}
\providecommand{\bibinfo}[2]{#2}
\providecommand{\BIBentrySTDinterwordspacing}{\spaceskip=0pt\relax}
\providecommand{\BIBentryALTinterwordstretchfactor}{4}
\providecommand{\BIBentryALTinterwordspacing}{\spaceskip=\fontdimen2\font plus
\BIBentryALTinterwordstretchfactor\fontdimen3\font minus
  \fontdimen4\font\relax}
\providecommand{\BIBforeignlanguage}[2]{{%
\expandafter\ifx\csname l@#1\endcsname\relax
\typeout{** WARNING: IEEEtran.bst: No hyphenation pattern has been}%
\typeout{** loaded for the language `#1'. Using the pattern for}%
\typeout{** the default language instead.}%
\else
\language=\csname l@#1\endcsname
\fi
#2}}
\providecommand{\BIBdecl}{\relax}
\BIBdecl

\bibitem{sw}
T.~F. Smith and M.~S. Waterman, ``Identification of common molecular
  subsequences,'' \emph{Journal of Mol. Biol.}, vol. 147, no.~1, pp. 195--197,
  1981.

\bibitem{nw}
S.~B. Needleman and C.~D. Wunsch, ``A general method applicable to the search
  for similarities in the amino acid sequence of two proteins,'' \emph{Journal
  of Mol. Biol.}, vol.~48, no.~3, pp. 443--453, 1970.

\bibitem{swaphils}
Y.~Liu, T.~T. Tran, F.~Lauenroth, and B.~Schmidt, ``{SWAPHI-LS:} smith-waterman
  algorithm on xeon phi coprocessors for long {DNA} sequences,'' in \emph{2014
  {IEEE} CLUSTER}, 2014, pp. 257--265.

\bibitem{hou2016aalign}
K.~Hou, H.~Wang, and W.-c. Feng, ``Aalign: A simd framework for pairwise
  sequence alignment on x86-based multi-and many-core processors,'' in
  \emph{2016 IEEE IPDPS}.\hskip 1em plus 0.5em minus 0.4em\relax IEEE, 2016,
  pp. 780--789.

\bibitem{misra2018performance}
S.~Misra, T.~C. Pan, K.~Mahadik, G.~Powley, P.~N. Vaidya, M.~Vasimuddin, and
  S.~Aluru, ``Performance extraction and suitability analysis of multi-and
  many-core architectures for next generation sequencing secondary analysis,''
  in \emph{Proc. of the 27th International Conference on Parallel Architectures
  and Compilation Techniques}.\hskip 1em plus 0.5em minus 0.4em\relax ACM,
  2018, p.~3.

\bibitem{cudalign}
E.~F.~O. Sandes and A.~C. de~Melo, ``Cudalign: using gpu to accelerate the
  comparison of megabase genomic sequences,'' in \emph{Proceedings PPoPP 2010},
  vol.~45, no.~5.\hskip 1em plus 0.5em minus 0.4em\relax ACM, 2010, pp.
  137--146.

\bibitem{liu2013cudasw++}
Y.~Liu, A.~Wirawan, and B.~Schmidt, ``Cudasw++ 3.0: accelerating smith-waterman
  protein database search by coupling cpu and gpu simd instructions,''
  \emph{BMC Bioinformatics}, vol.~14, no.~1, p. 117, 2013.

\bibitem{korpar2013sw}
M.~Korpar and M.~{\v{S}}iki{\'c}, ``Sw\#--gpu-enabled exact alignments on
  genome scale,'' \emph{Bioinformatics}, vol.~29, no.~19, pp. 2494--2495, 2013.

\bibitem{de2016cudalign}
E.~F. de~Oliveira~Sandes, G.~Miranda, X.~Martorell, E.~Ayguade, G.~Teodoro, and
  A.~C.~M. Melo, ``Cudalign 4.0: Incremental speculative traceback for exact
  chromosome-wide alignment in gpu clusters,'' \emph{IEEE TPDS}, vol.~27,
  no.~10, pp. 2838--2850, 2016.

\bibitem{oliver2005hyper}
T.~Oliver, B.~Schmidt, and D.~Maskell, ``Hyper customized processors for
  bio-sequence database scanning on fpgas,'' in \emph{Proceedings of the 2005
  ACM/SIGDA FPGA}.\hskip 1em plus 0.5em minus 0.4em\relax ACM, 2005, pp.
  229--237.

\bibitem{fpga}
I.~Li, W.~Shum, and K.~Truong, ``160-fold acceleration of the smith-waterman
  algorithm using a field programmable gate array (fpga),'' \emph{BMC
  Bioinformatics}, vol.~8, 2007.

\bibitem{rucci2018swifold}
E.~Rucci, C.~Garcia, G.~Botella, A.~De~Giusti, M.~Naiouf, and M.~Prieto-Matias,
  ``Swifold: Smith-waterman implementation on fpga with opencl for long dna
  sequences,'' \emph{BMC Systems Biology}, vol.~12, no.~5, p.~96, 2018.

\bibitem{reinert2017seqan}
K.~Reinert, T.~H. Dadi, M.~Ehrhardt, H.~Hauswedell, S.~Mehringer, R.~Rahn,
  J.~Kim, C.~Pockrandt, J.~Winkler, E.~Siragusa \emph{et~al.}, ``The seqan c++
  template library for efficient sequence analysis: A resource for
  programmers,'' \emph{Journal of Biotechnology}, vol. 261, pp. 157--168, 2017.

\bibitem{parasail}
J.~Daily, ``Parasail: Simd c library for global, semi-global, and local
  pairwise sequence alignments,'' \emph{BMC Bioinformatics}, vol.~17, no.~1,
  p.~1, 2016.

\bibitem{ssw}
M.~Zhao, W.-P. Lee, E.~P. Garrison, and G.~T. Marth, ``Ssw library: an simd
  smith-waterman c/c++ library for use in genomic applications,'' \emph{PloS
  one}, vol.~8, no.~12, p. e82138, 2013.

\bibitem{nvbio}
J.~Pantaleoni and N.~Subtil, ``{NVBIO},'' \url{https://nvlabs.github.io/nvbio},
  2015.

\bibitem{DBLP:conf/gpce/LeissaBHMS15}
R.~Lei{\ss}a, K.~Boesche, S.~Hack, R.~Membarth, and P.~Slusallek, ``Shallow
  embedding of dsls via online partial evaluation,'' in \emph{Proc. of the 2015
  {ACM} {SIGPLAN} International Conference on Generative Programming: Concepts
  and Experiences, {GPCE} 2015}, 2015, pp. 11--20.

\bibitem{DBLP:journals/pacmpl/LeissaBHPMSMS18}
R.~Lei{\ss}a, K.~Boesche, S.~Hack, A.~P{\'{e}}rard{-}Gayot, R.~Membarth,
  P.~Slusallek, A.~M{\"{u}}ller, and B.~Schmidt, ``{AnyDSL}: A partial
  evaluation framework for programming high-performance libraries,''
  \emph{{PACMPL}}, vol.~2, no. {OOPSLA}, pp. 119:1--119:30, 2018.

\bibitem{Futamura:1999:PEC:609149.609205}
Y.~Futamura, ``Partial evaluation of computation process--an approach to a
  compiler-compiler,'' \emph{Higher-Order and Symbolic Computation}, vol.~12,
  no.~4, pp. 381--391, Dec 1999, revision of the 1971 paper.

\bibitem{Consel}
U.~P. Schultz, J.~L. Lawall, and C.~Consel, ``Automatic program specialization
  for java,'' \emph{{ACM} Trans. Program. Lang. Syst.}, vol.~25, no.~4, pp.
  452--499, 2003.

\bibitem{Brady}
E.~Brady and K.~Hammond, ``Scrapping your inefficient engine: using partial
  evaluation to improve domain-specific language implementation,'' in
  \emph{Proceeding of the 15th {ACM} {SIGPLAN} international conference on
  Functional programming, {ICFP} 2010}, 2010, pp. 297--308.

\bibitem{DBLP:conf/rims/Futamura82}
Y.~Futamura, ``Parital computation of programs,'' in \emph{Proceedings of the
  {RIMS} Symposia on Software Science and Engineering}, 1982, pp. 1--35.

\bibitem{DBLP:conf/esop/Consel88}
C.~Consel, ``New insights into partial evaluation: the {SCHISM} experiment,''
  in \emph{{ESOP} '88, 2nd European Symposium on Programming, Nancy, France,
  March 21-24, 1988, Proceedings}, 1988, pp. 236--246.

\bibitem{hirschberg75}
D.~S. Hirschberg, ``A linear space algorithm for computing maximal common
  subsequences,'' \emph{Comm. ACM}, vol.~18, 1975.

\bibitem{DBLP:conf/pldi/MollH18}
S.~Moll and S.~Hack, ``Partial control-flow linearization,'' in
  \emph{Proceedings of the 39th {ACM} {SIGPLAN} Conference on Programming
  Language Design and Implementation, {PLDI} 2018}, 2018, pp. 543--556.

\bibitem{rahn2018generic}
R.~Rahn, S.~Budach, P.~Costanza, M.~Ehrhardt, J.~Hancox, and K.~Reinert,
  ``Generic accelerated sequence alignment in seqan using vectorization and
  multi-threading,'' \emph{Bioinformatics}, vol.~34, no.~20, pp. 3437--3445,
  2018.

\bibitem{wozniak1997using}
A.~Wozniak, ``Using video-oriented instructions to speed up sequence
  comparison,'' \emph{Bioinformatics}, vol.~13, no.~2, pp. 145--150, 1997.

\bibitem{farrar2006striped}
M.~Farrar, ``Striped smith--waterman speeds database searches six times over
  other simd implementations,'' \emph{Bioinformatics}, vol.~23, no.~2, pp.
  156--161, 2006.

\bibitem{Maleki}
S.~Maleki, M.~Musuvathi, and T.~Mytkowicz, ``Low-rank methods for parallelizing
  dynamic programming algorithms,'' \emph{{TOPC}}, vol.~2, no.~4, pp.
  26:1--26:32, 2016.

\end{thebibliography}

\end{document}